# Atomically flat single-crystalline gold nanostructures for plasmonic nanocircuitry

Jer-Shing Huang[1,*,§], Victor Callegari[2], Peter Geisler[1], Christoph Brüning[1], Johannes Kern[1], Jord C. Prangsma[1], Xiaofei Wu[1], Thorsten Feichtner[1], Johannes Ziegler[1], Pia Weinmann[3], Martin Kamp[3], Alfred Forchel[3], Paolo Biagioni[4], Urs Sennhauser[2] & Bert Hecht[1,†]

[1]*Nano-Optics & Biophotonics Group, Experimentelle Physik 5, Physikalisches Institut, Wilhelm-Conrad-Röntgen-Center for Complex Material Systems, Universität Würzburg, Am Hubland, D-97074 Würzburg, Germany*

[2]*EMPA, Swiss Federal Laboratories for Materials Testing and Research, Electronics/Metrology/Reliability Laboratory, Ueberlandstrasse 129, CH-8600 Dübendorf, Switzerland*

[3]*Technische Physik, Physikalisches Institut, Wilhelm-Conrad-Röntgen-Center for Complex Material Systems, Universität Würzburg, Am Hubland, D-97074 Würzburg, Germany*

[4]*CNISM - Dipartimento di Fisica, Politecnico di Milano, Piazza Leonardo da Vinci 32, 20133 Milano, Italy*

\* To whom correspondence should be addressed: jshuang@mx.nthu.edu.tw

† To whom correspondence should be addressed: hecht@physik.uni-wuerzburg.de

§ J.-S.H. current address:
Department of Chemistry, National Tsing Hua University, Hsinchu 30013, Taiwan

**Deep subwavelength integration of high-definition plasmonic nanostructures is of key importance for the development of future optical nanocircuitry for high-speed communication, quantum computation, and lab-on-a-chip applications. So far the experimental realization of proposed extended plasmonic networks consisting of multiple functional elements remains challenging, mainly due to the multi-crystallinity of commonly used thermally evaporated gold layers. Resulting structural imperfections in individual circuit elements will drastically reduce the yield of functional integrated nanocircuits. Here we demonstrate the use of very large (>100 $\mu m^2$) but thin (<80 nm) chemically grown single-crystalline gold flakes, which, after immobilization, serve as an ideal basis for focused-ion beam milling and other top-down nanofabrication techniques on any desired substrate. Using this methodology we obtain high-definition ultrasmooth gold nanostructures with superior optical properties and reproducible nano-sized features over micrometer length scales. Our approach provides a possible solution to overcome the current fabrication bottleneck and to realize high-definition plasmonic nanocircuitry.**

The recent years have seen tremendous activity in the development of basic building blocks of optical nanocircuitry and improved photovoltaic devices which take advantage of plasmonic resonances of noble metals and the associated strongly enhanced local fields[1-4]. Sub-wavelength plasmonic waveguides[5-6], optical nanoantennas[7-8], as well as various plasmonic modulators[9-10] and resonators for ultra sensing and microscopy[8,11-14] have been suggested and realized experimentally. Furthermore, the strongly enhanced local fields associated with plasmon resonances have been exploited to boost various nonlinear optical phenomena[15-16]. Recently, first steps have been taken to transfer concepts of quantum optics to plasmonics[17-20]. In combination with coherent control techniques for near-field manipulation[21-23], gain materials for field amplification[24-27] and the impedance matching concept for building up optical nanocircuitry[28-29], functional plasmonic nanocircuitry operating at optical frequencies is about to become a field of major scientific and technological impact.

However, when it comes to advanced functional - and therefore necessarily more complex - plasmonic nanostructures, theoretical studies using numerical simulations are far ahead of what is currently in reach of state-of-the-art micro- and nanofabrication techniques [see for example refs. 22, 28, 29]. This trend roots in the very small fabrication tolerances that are necessary to yield a desired functionality. Small enough tolerances, however, are very difficult to obtain due to the multi-crystalline structure of thin gold films produced by thermal evaporation[30]. As an illustration we consider a plasmonic transmission line consisting of two wires separated by a nanometer sized gap extending of micrometer distances. While such transmission lines perform well in simulations, in a real structure, fabricated by state-of-the-art nanofabrication techniques, already a single nanometer-scale defect in the gap can lead to a strong power reflection due to the local impedance change and render the structure useless. As a general rule, fabrication tolerances become more critical as the degree of field confinement and enhancement in plasmonic nanostructures increases. For isolated nanostructures with a single critical dimension, such as the feedgap of a nanoantenna, insufficient fabrication tolerances can be compensated by producing large arrays of similar structures and selecting individuals that match the specification. This approach, however, breaks down as soon as several nanostructures, each with their own critical dimensions, need to be combined in a complex device because the yield of functional devices then decreases rapidly with the number of elements. In addition there is clear evidence that in multi-crystalline nanostructures scattering of plasmons is enhanced[30-32], which has negative consequences for both propagation effects and the achievable maximal near-field intensity enhancement. For the progress of nanoplasmonics, it is, therefore crucial to have methods at hand that allow precise fabrication of complex, functional, single-crystalline plasmonic nanostructures and extended networks.

When applied to the task of manufacturing complex plasmonic nanostructures, electron beam lithography (EBL), the currently most popular fabrication method, suffers from the multi-crystalline character of gold layers and ensuing imperfections introduced by the lift-off process[30,33]. Moreover, in order to enhance the sticking of very small gold nanostructures to various substrates, including ITO-coated substrates providing DC conductivity, a standard method is to deposit a thin adhesion layer, such as 5 nm of titanium or chromium. This adhesion layer, unfortunately, damps the local plasmonic resonance drastically and, therefore, reduces the plasmon propagation length and near-field intensity enhancement[34-37]. To avoid the adhesion layer, Mühlschlegel et al.[7] have chosen an approach combining EBL and focused-ion beam (FIB) milling to fabricate optical nanoantennas. They first use EBL to fabricate marker structures and large gold patches (>1 µm$^2$), which, during the lift-off, stay firmly attached to the substrate even without adhesion layer. Subsequent FIB milling then yields the desired structures. Using this approach, white-light continuum generation with very low excitation powers (down to 20 µW) has been observed for some of the antennas nominally in resonance with the excitation. Since the gold film produced by vapour deposition consists of randomly oriented crystal grains, the precision of the fabricated structures is limited by the size of the grains due to the fact that different crystal domains show different resistance to FIB milling[33]. The typical diameter of crystal grains in thin layers of vapour-deposited gold is about 30 nm to 50 nm (see Figure 1c), which impedes the fabrication of structures containing features of comparable size and introduces an intrinsic surface roughness for larger structures which leads to scattering[31] and increased dephasing of surface plasmons. Alternative fabrication methods such as template stripping employing a patterned silicon substrate[38] or induced-deposition mask lithography (IDML)[14] retain the problem of multicrystallinity and therefore cannot effectively remove the nanofabrication bottleneck in plasmonics.

As a new paradigm for the fabrication of extended functional plasmonic nanostructures, we propose using chemically synthesized single-crystalline gold flakes[39-41](Figure 1a) which are deposited on a substrate and subsequently structured by FIB milling. Using this combination of bottom-up and top-down nanofabrication, we have obtained isolated nanoantennas and more complex plasmonic nanostructures for optical nanocircuitry with superior optical quality, well-defined dimensions and crystallographic orientation, as well as atomically flat surfaces. The use of chemically synthesized single-crystalline metal flakes is inexpensive and not limited to gold[42]. However the chemical stability and degradation at ambient conditions may need to be taken into consideration for different metals. It can be applied to all kinds of substrates and thus may facilitate hybrid plasmonic waveguiding[6] and plasmonic lasing[27], where a well-controlled deep sub-wavelength contact between different materials is necessary. Atomically flat structures may also prove beneficial for high-precision measurements as in the area of Casimir interaction[43-45] and plasmonic optical trapping[46-49].

**Results**

**Properties of single crystalline gold flakes.** The growth of large gold flakes is achieved following the procedure described in ref. 40 (see Methods). As shown in Figure 1a, the single-crystalline flakes appear as triangles and truncated triangles. The high surface quality of single-crystalline flakes (Figure 1b) is confirmed by both very low surface roughness (<1 nm) over a large area (1 $\mu m^2$) as determined by atomic force microscopy (AFM, Figure S4, Supplementary Information) and the complete absence of the TPPL signal (Figure S5, Supplementary Information), which is expected to be strongly enhanced in presence of surface roughness[50], as normally seen in a multi-crystalline metal film (Figure 1c).

The thickness varies from flake to flake between 40 nm and 80 nm but is constant within one flake. Such constant thickness, together with the single crystallinity, allows for reproducible fabrication of complete functional plasmonic circuits or arrays of isolated structures with critical dimensions over large areas using a constant and optimized ion beam condition within a single flake. To our experience, clean gold flakes stick sufficiently well to the substrate spontaneously. Nanostructures fabricated from such flakes are robust during transportation and optical as well as electron beam and AFM investigation of the sample despite the absence of an adhesion layer in accordance with previous experiments on gold nanostructures without adhesion layer[7]. We note that if necessary, sticking may, however, be improved by introducing covalent molecular linkage between suitably functionalized flakes and surfaces.

**Focused ion-beam milling of single crystalline gold flakes.** As a result, we are able to fabricate nanostructures that exhibit ultrasmooth surfaces and small gaps over extended distances, which is not possible using any other approach. Figure 1d shows a high resolution transmission electron microscopy (TEM) image recorded at the FIB-milled edge of a single-crystalline flake on a carbon film substrate. This image indicates that the crystallinity of the gold flakes is hardly affected by the ion beam. However, surface contamination with very small (1-2 nm) particles possibly due to re-deposition of the sputtered material are observed. Another anticipated intrinsic side-effect of FIB milling is the Ga ion implantation. We have performed energy dispersive X-ray (EDX) analysis of gold flakes at a position close to a FIB-milled edge and the result shows no measurable Ga ion implantation, corresponding to a Ga ion concentration of less than 1% (see Figure S9, Supplementary Information). The effects of Ga ion implantation are not clear and need further investigation. Nevertheless, our method can be combined with various alternative top-down approaches, such as IDML[14], to avoid such possible side effects.

In Figure 2 we compare the quality of nanostructures fabricated from single-crystalline gold flakes (right) and from vapour-deposited multi-crystalline gold layers of the same thickness (left) fabricated using the same FIB milling pattern and optimized conditions. It is evident that nanostructures with very fine features that extend over large areas can be easily fabricated in single-crystalline gold flakes. Although for multi-crystalline structures similar structural details (e.g. wire separation) can be achieved, the presence of grains leads to unpredictable structural defects, such as particles bridging the gap in Figure 2a and Figure 2b. The high quality of the structures is reproducible and greatly improved compared to previously used structures [see for example refs. 8, 15, 30 & 38]. In contrast to structures fabricated using vapour-deposited multi-crystalline gold films, the quality of plasmonic nanostructures obtained by our new method is independent of the surface roughness of the ITO substrate (compare Figure 2a and Figure 2b for the effect of different ITO substrates).

**Optical properties of single crystalline gold nanostructures.** In the following we show that our method not only improves the structural quality but also the optical properties by reducing the scattering of surface plasmons, similar to improvements observed for single-crystalline silver wires[32] and gold structures[31,51]. Figure 3a shows a scanning electron microscope (SEM) image of bowtie nanoantennas fabricated side-by-side on a single-crystalline gold flake and on a multi-crystalline gold patch prepared by EBL. Note that a 5 nm thick titanium adhesion layer is used here for the multi-crystalline gold film representing the commonly used metal film in EBL.

We probe the quality of plasmonic nanostructures by recording their two-photon excited photoluminescence (TPPL, see Methods). TPPL has been extensively used for the characterization of the resonant behaviour of nanoparticles[7,50,52-55]. The TPPL intensity is proportional to $I^2$, where $I$ is the local near-field intensity enhancement[50,52-53]. The recorded TPPL map (Figure 3b) shows that in addition to the distinct improvement of the structural quality, the single-crystalline antennas show a much higher TPPL emission count rate (>100 times) upon resonant excitation, indicating much larger field enhancement. The much lower TPPL signal of the multi-crystalline antennas is attributed to the increased damping and slight resonance shift due to the adhesion layer as well as due to enhanced plasmon scattering in the multi-crystalline gold[31]. Besides, since typical materials used for adhesion layers, i.e. Ti and Cr, have higher resistance to FIB milling[33], the minimal ion dose optimized for finest spatial resolution can be insufficient to completely remove the adhesion layer in the gap. Consequently, the impedance and optical properties of the nanoantennas can be strongly influenced due to the remaining material in the gap, which acts as a nanoload[28,56]. For the multi-crystalline bowtie antennas in Figure 3, the excitation power needs to be increased up to 170 µW to obtain a measurable TPPL signal as compared to <10 µW for the single-crystalline ones. This may be attributed to the fact that, according to numerical simulations, the source excitation spectrum (centered at 828 nm) does not exactly hit but only overlaps partially with the broad resonance peak of the bowtie antennas (Figure S8, Supplementary Information). Any scattering or damping of the near-field reduces the TPPL signal drastically due to the quadratic dependence of TPPL on the integrated intensity in the gold structures.

**Effects of the adhesion layer.** In order to disentangle the damping introduced by the adhesion layer and the damping due to increased scattering at crystal domain boundaries, we have also prepared multi-crystalline gold patches without adhesion layer using EBL. To allow for a direct comparison of single- and multi-crystalline structures, the thickness of the gold flake is reduced to 30 nm using a low energy ion beam before writing the nanoantennas, such that the height of single- and multi-crystalline nanoantennas is nearly identical (see Methods), as confirmed by AFM (Figure S2, supplementary Information). We have also acquired the white-light scattering spectra for both single- and multi-crystalline antenna arrays, which show very close peak positions (Figure S10, supplementary Information). We therefore conclude that the structures that are compared have nominally the same structural dimensions and resonance frequencies. The difference in TPPL intensity can therefore be attributed to differences in the local field enhancement of the structures. The SEM image (Figure 4a) shows linear dipole antenna arrays fabricated using both a single-crystalline gold flake and the multi-crystalline gold film (marked with white-dashed rectangle, 4 rows each containing antennas with the same dimensions). Upon longitudinally polarized illumination (828 nm, 1 ps, 50 µW), both single- and multi-crystalline antenna arrays now exhibit easily detectable TPPL emission. Furthermore, due to the fact that the antenna length is increasing in steps of 20 nm, from left to right in the arrays (196 nm to 396 nm, respectively), a dependence of the TPPL emission rate on the antenna length is observed (Figure 4b-c), which reaches a maximum for those antennas whose resonance has the best spectral overlap with the excitation source. A second maximum and the spot shape transformation related to the antibonding mode[52] are also seen. We observe that on average, the single-crystalline antennas show much higher emission count rates even though the structural parameters of most single- and multi-crystalline antenna pairs are nearly identical (Figure S2, Supplementary Information).

In addition to the higher count rate, the single-crystalline antenna array much better reproduces the length-dependent intensity evolution as predicted by FDTD simulations. In particular, we note that some of the multi-crystalline antennas remain completely dark (see e.g. the yellow-dashed circle in Figure 4). Checking the respective SEM images, it is found that all dark antennas have gaps not completely cut through, possibly due to the presence of grains. It is further observed that overall the structural quality of single-crystalline antennas is significantly improved compared to the multi-crystalline counterparts. While many of the multi-crystalline antennas exhibit structural defects (see Figure 4d, upper row), including defective gaps, the single-crystalline antennas very reproducibly display homogenous structural properties, such as width, gap, shape and relative orientation of the two antenna arms. In particular, the examples shown in Figure 4 clearly illustrate our finding that dipole antennas fabricated from single-crystalline flakes reproducibly show the predicted length-dependent behaviour since the gap width can be kept constant over a whole array of structures. In contrast, for a multi-crystalline gold film without adhesion layer the gap width shows strong variations - only occasionally structures of high quality can be found.

## Discussion

Compared to single-crystalline gold layers prepared by the Czochralski process and subsequent polishing[31,51] or the epitaxial growths of gold layers on lattice-matched substrates, such as mica or MgO, the method proposed here using chemically synthesized gold flakes is easier, cheaper, and requires no specialized instrumentation while the freedom of choosing a substrate at will is retained. The obtained metal layers have well-defined crystal orientation and very large aspect ratio. The large flake area, single crystallinity, and constant thickness within one flake facilitate reproducible top-down fabrication and render precise positioning of the flake on a substrate unnecessary. However, promising nanoparticle alignment techniques[57,58] and nanomanipulation[59] have been reported and may be applied to gold flakes as well. In addition, since the crystal orientation can be directly identified from the shape of the flakes, nanostructures can be easily fabricated with distinct crystal orientations determined by the orientation of the FIB milling pattern with respect to the single-crystalline gold flake. This would allow a direct measurement of the optical properties of nanostructures as a function of the crystal orientation.

In conclusion we demonstrate a new method for the FIB fabrication of single-crystalline plasmonic nanostructures using large and thin, chemically grown single-crystalline gold flakes deposited onto a glass/ITO surface. This combination of bottom-up and top-down nanofabrication yields greatly improved fabrication tolerances as well as improved structural homogeneity as compared to conventional multi-crystalline structures. We are therefore able to fabricate plasmonic gold nanostructures with reproducible and well-defined nanometer scale features extending over micrometer length scales. We demonstrate the improved optical quality of our structures by means of the enhanced TPPL emission of single-crystalline linear dipole and bowtie optical antennas which serves as a benchmark for their strong near-field intensity enhancement. Our method provides possible solutions for the fabrication and realization of high-definition complex plasmonic nanodevices and extended optical nanocircuits, as well for precise measurement of nanoscale strong interaction forces.

## Methods

**Flake fabrication & sample preparation.** The growth of large gold flakes is achieved following the procedure described in ref. 40. The reaction temperature has been reduced to 60˚C (Figure S3, Supplementary Information) to obtain thin flakes (thickness < 80 nm) with large surface area (> 100 μm$^2$). The produced gold flake suspension is rinsed and stored as aqueous suspension. We note that further processing of the rinsed suspension, such as centrifuging or filtering may help to remove unwanted small particles but also decreases the yield of the flakes. 5 minutes ultrasonication is used to homogeneously disperse the gold flakes before directly drop-casting the suspension onto ITO coated cover glasses, on which multi-crystalline gold marker-structures with 30 nm thickness are pre-fabricated by EBL. The marker structures are designed to be easily observable in both conventional optical microscopy and SEM and therefore to facilitate the precise localization and identification of the same gold flakes using both methods. In addition, vapour-deposited markers provide multi-crystalline gold films for control experiments. Isolated gold flakes of sufficiently large size and spatially homogeneous optical transmission are pre-selected by optical microscopy. Selected flakes are then imaged by SEM to confirm the homogenous contact to the substrate and the absence of defects.

**FIB milling.** FIB milling (Helios Nanolab, FEI Company) is applied to single-crystalline flakes that pass all those criteria and are found nearby multi-crystalline marker structures serving as a reference. Best structuring results are obtained for 30 kV acceleration voltage and 1.5 pA Ga-ion current. The distance between adjacent structures is larger than 700 nm to minimize crosstalk and to ensure that only one structure is illuminated by the focal spot (FWHM = 350 nm). Using these settings, writing arrays of linear nanoantennas/bowties covering an area of 10 µm x 10 µm usually takes less than one hour. Before starting the antenna fabrication, the flake was thinned down to match the thickness of the multi-crystalline marker structure (30 nm). For fabrications of both single- and multi-crystalline antennas, FIB milling removes 20-25 nm of ITO layer and result in a topography contrast of about 55-60 nm from the milled ITO surface to the antenna upper surface (Figure S2, supplementary Information). Such geometrical details are taken into consideration in the numerical simulations. EDX analysis of gold flakes recorded at a FIB-milled edge shows no measurable Ga-ion implantation which corresponds to a Ga-ion concentration of less than 1% (see Figure S9, Supplementary Information).

**Two-photon photoluminescence imaging.** To obtain a TPPL intensity map, the ultrashort pulses from a mode-locked Ti:sapphire laser (center wavelength = 828 nm, 80 fs, 80 MHz, Time-Bandwidth Products, Tiger) are coupled into 1.5 m of optical fiber to get stretched to approximately 1 ps. The linear polarized fiber output is then collimated, passes a dichroic mirror (DCXR770, Chroma Technology Inc.) and is focused through the cover glass onto the nanostructures using an oil immersion microscope objective (Plan-Apochromat 100x, Oil, N.A. = 1.4, Nikon). The photoluminescence signal is collected by the same objective and is reflected by the dichroic mirror. Laser scattering and possible second harmonic signals are rejected by a holographic notch filter (O.D. > 6.0 at 830 nm, Kaiser Optical System, Inc.) and a bandpass filter (transmission window: 450-750 nm, D600/300, Chroma Technology Inc.) in front of the photon detector (SPCM-AQR 14, Perkin-Elmer). (see Supplementary Information for experiment details).

**Atomic force microscopy.** AFM measurement are carried out in ambient condition with tapping mode operating at a resonance frequency of 240-280 kHz and a scanning rate of 0.2 Hz (DMLS scanning head, Nanoscope IIIa, Digital instruments).

**White-light scattering spectroscopy.** To obtain white-light scattering spectra, an off-axis and s-polarized needle-like beam (collimated beam with 1-2 mm diameter) obtained from a halogen lamp (Axiovert, Zeiss) with polarization along the long axis of the antenna is focused by a microscope objective (Plan-Apochromat, 63×, N.A. = 1.4, Zeiss) to the sample plane. Since the off-axis beam is parallel to but displaced from the objectives optical axis such that it hits the surface at an angle larger than the critical angle, it undergoes total internal reflection at the sample plane. Nanoantennas that are illuminated scatter light into a broad angular range collected by the same objective while the reflected excitation beam is blocked by a small beam stop in the detection path. An achromatic λ/2 waveplate and a polarizer (Glan-Thompson calcite polarizer) are used to select the polarization of the scattered light before the entrance slit (200 µm) of the spectrometer (ACTON SpectraPro 2300i, 150 grooves/mm grating blazing at 500 nm). The typical acquisition time of the CCD is 20 seconds. All spectra are normalized with the source spectrum and the overall transmission efficiency of the optical system.

**Numerical simulations.** Numerical simulations adopting the nominal antenna dimensions used in FIB milling are performed with a commercial finite-difference time-domain solver (FDTD Solutions v6.5.8, Lumerical Solutions, Inc.). The dielectric constant of gold is modeled according to experimental data[60]. The minimal mesh size is set to 1 nm$^3$. Structures are made of gold and placed on top of an ITO layer (thickness = 100 nm). All boundaries of the simulation box are set to be at least 700 nm away from the structure to avoid spurious absorption of the near fields. The geometry changes of the substrate due to the FIB milling are also taken into account. To mimic the experimental conditions, the source is set to have the same spectral width as the laser used in the experiment (821 - 835 nm) and is focused onto the gold/ITO interface using a thin lens (N.A. = 1.4) approximated by a superposition of 200 plane waves. The total electric field intensity ($|E_x|^2 + |E_y|^2 + |E_z|^2$) distribution inside both antenna arms is recorded using 3D field profile monitors. A quantity proportional to the TPPL signal is obtained by integrating the square of the field intensity ($\iiint \{|E_x|^2 + |E_y|^2 + |E_z|^2\}^2 dxdydz$) over the volume of the antenna arms[52].

## Acknowledgement
The authors thank Prof. N. Gu and M. Mitterer for valuable discussions and assistance in the synthesis of gold flakes. We also thank S. Meier, T. Schmeiler and M. Emmerling for the assistance in FIB, AFM and EBL, respectively. Financial support of the DFG via the SPP1391 is gratefully acknowledged.

## Author contributions
B.H. conceived the original idea. J.-S.H. and C.B. synthesized the gold flakes. J.-S.H. and V.C. performed FIB and SEM. P.W. performed EBL. M.K. and J.C.P. performed TEM and EDX. J.C.P and T.F. performed AFM. J.-S.H., P.G., J.K., X.W. and J.Z. carried out the optical experiments. J.-S.H. designed and implemented numerical simulations; analyzed and assembled the data. J.-S.H. and B.H. wrote the manuscript. All authors contributed to scientific discussions and critical revision of the article. M.K., A.F., U.S. and B.H. supervised the study.

## Competing financial interests
The authors declare no competing financial interests.

## Additional information
Supplementary information accompanies this paper at www.nature.com/ncommunications.
Reprints and permission information is available online at http://npg.nature.com/reprintsandpermissions/.
Correspondence and requests for materials should be addressed to
J.-S.H. (jshuang@mx.nthu.edu.tw) or B.H. (hecht@physik.uni-wuerzburg.de)

# Figure Captions

**Figure 1 | SEM and TEM images of chemically synthesized gold flakes and vapour-deposited gold film.** (a) Overview of a cluster of self-assembled single-crystalline gold flakes. The thickness of the flakes usually varies between 40 nm and 80 nm, but is homogeneous for each flake. (b) Zoomed-in SEM image of the surface of a single-crystalline flake with rectangular area milled by FIB. The nearby particle is a nanoparticle from the suspension. (c) Zoomed-in SEM image of a typical surface of vapor-deposited multi-crystalline gold film (marker structure) consisting of randomly orientated grains (30 – 50 nm) on top of an ITO substrate. Note that (b) and (c) are recorded with 52˚ tilt angle. (d) High-resolution TEM image of a single-crystalline gold flake with the edge created by FIB milling. The single-crystal domain remains after FIB milling with small particles (1-2 nm) due to re-deposition of the sputtered material (black arrow).

**Figure 2 | SEM images of single- and multi-crystalline gold nanostructures.** (a) Prototype optical nanocircuits (refs. 22,29) fabricated using FIB on a vapor-deposited multi-crystalline (left) and a single-crystalline gold flake deposited on a dip-coated ITO substrate. The length of nanoantennas and the two-wire transmission line are 380 nm and 4 µm, respectively. Note that images in (a) are recorded with 52˚ tilt angle. (b) Optical nanocircuits with a 90˚ corner fabricated with FIB on multi-crystalline gold film (inset) and a single-crystalline gold flake on top of sputtered ITO substrate. (c) Asymmetric bull's eye (annular resonator) nanostructure fabricated with FIB on single- (right) and multi-crystalline (left) gold film. All scale bars are 500 nm. Note the increased roughness and structural imperfections of the multi-crystalline structures.

**Figure 3 | SEM image and corresponding TPPL map of single-crystalline and multi-crystalline gold bowtie nanoantennas** (a) SEM image of bowtie antennas fabricated by FIB using a self-assembled single-crystalline gold flake and a vapor-deposited multi-crystalline gold film with 5 nm titanium adhesion layer on top of a sputtered ITO substrate. (b) Map of visible TPPL of the same area shown in (a), obtained by scanning the sample over the tightly focused laser spot ($\lambda$ = 828 nm, N. A. = 1.4, average power = 70 µW, pulse duration = 1 ps ) and recording the emission intensity with a notch filter (O.D. > 6 at 830 nm) and a bandpass filter (frequency window= 450 - 750 nm).

**Figure 4 | SEM image and corresponding TPPL map of single-crystalline and multi-crystalline linear dipole nanoantennas** (a) SEM overview image of the entire area subject to FIB milling showing a large single-crystalline gold flake and patches of a vapor-deposited multi-crystalline gold film without adhesion layer on top of a sputtered ITO substrate. (b) TPPL map of the area marked with the white-dashed rectangle in (a). (c) Averaged integrated TPPL intensity obtained from nanoantennas on single-crystalline (red dots) and multi-crystalline (black triangles) gold film in the white-dashed area in (a), plotted versus nominal antenna length together with results obtained from the simulation (blue open squares, blue solid line is a guide for the eye). All experimental results are normalized to the maximal signal obtained from single-crystalline nanoantennas and the error bars indicate the standard deviation obtained from measurements on four nominally identical but distinct antenna arrays on the same flake. (d) Zoomed-in SEM images of the antenna series indicated by the white arrows in (a).

**Figures**
Figure 1    J.-S. Huang *et al.*

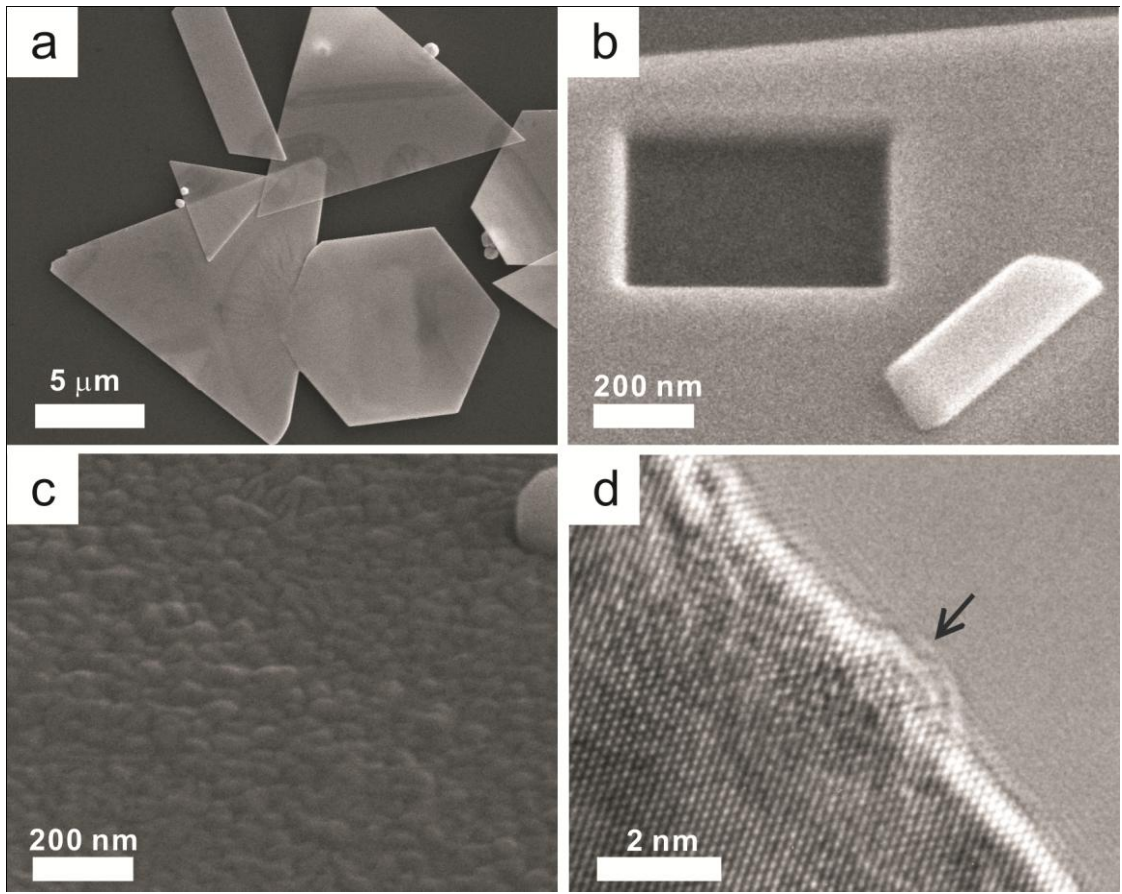

Figure 1    J.-S. Huang *et al.*

Figure 2     J.-S. Huang *et al.*

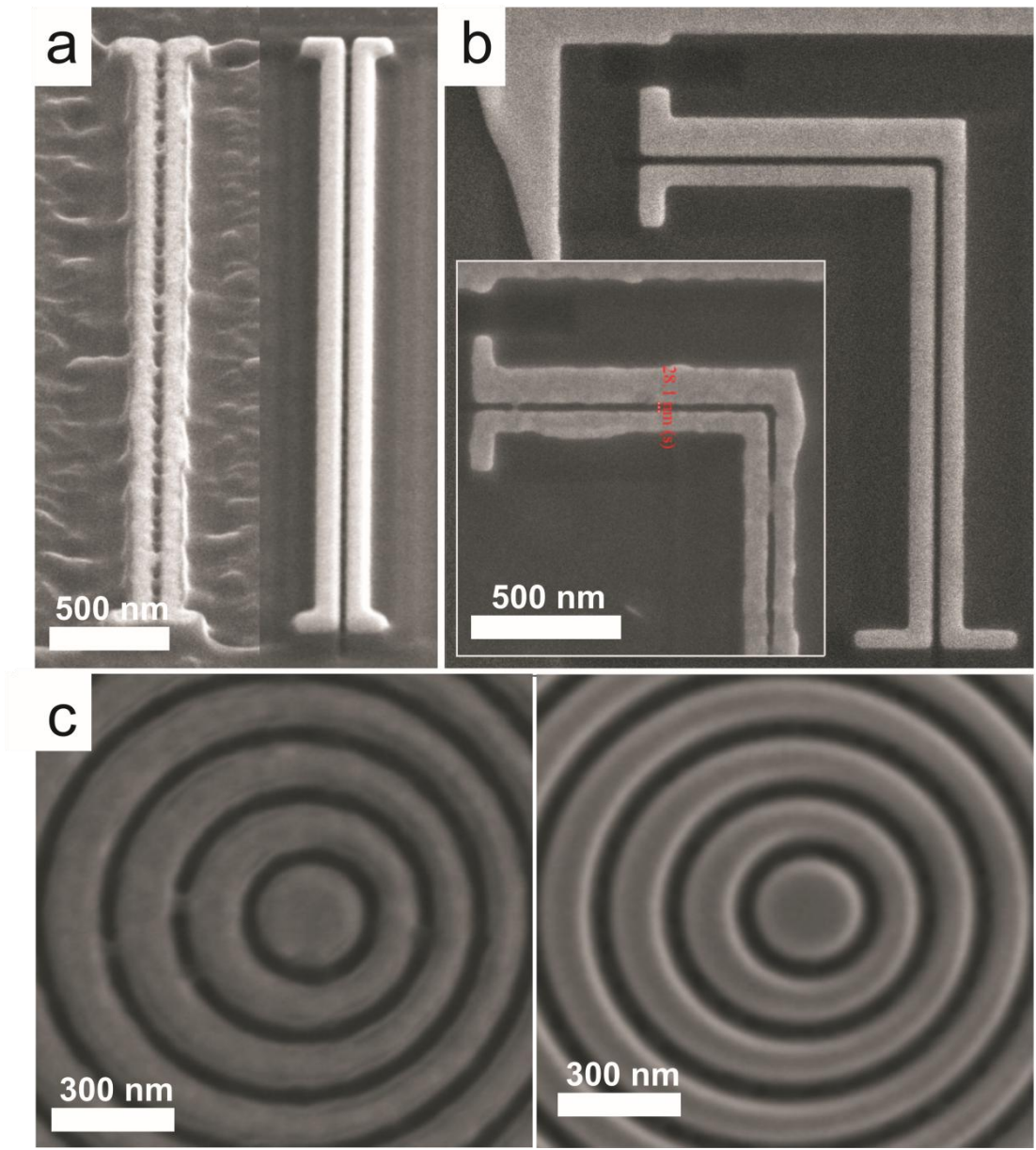

Figure 3     J.-S. Huang *et al.*

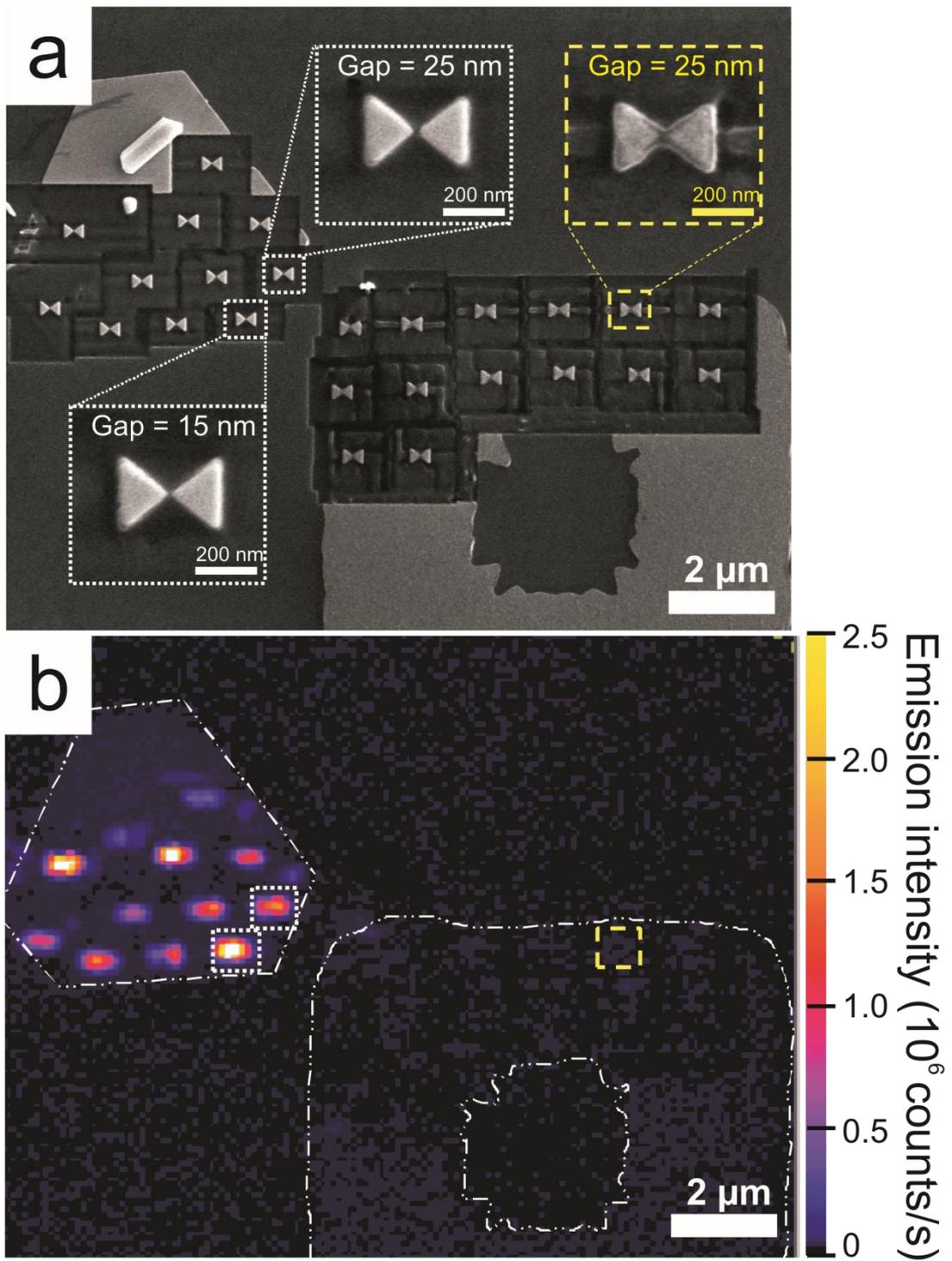

Figure 4    J.-S. Huang *et al.*

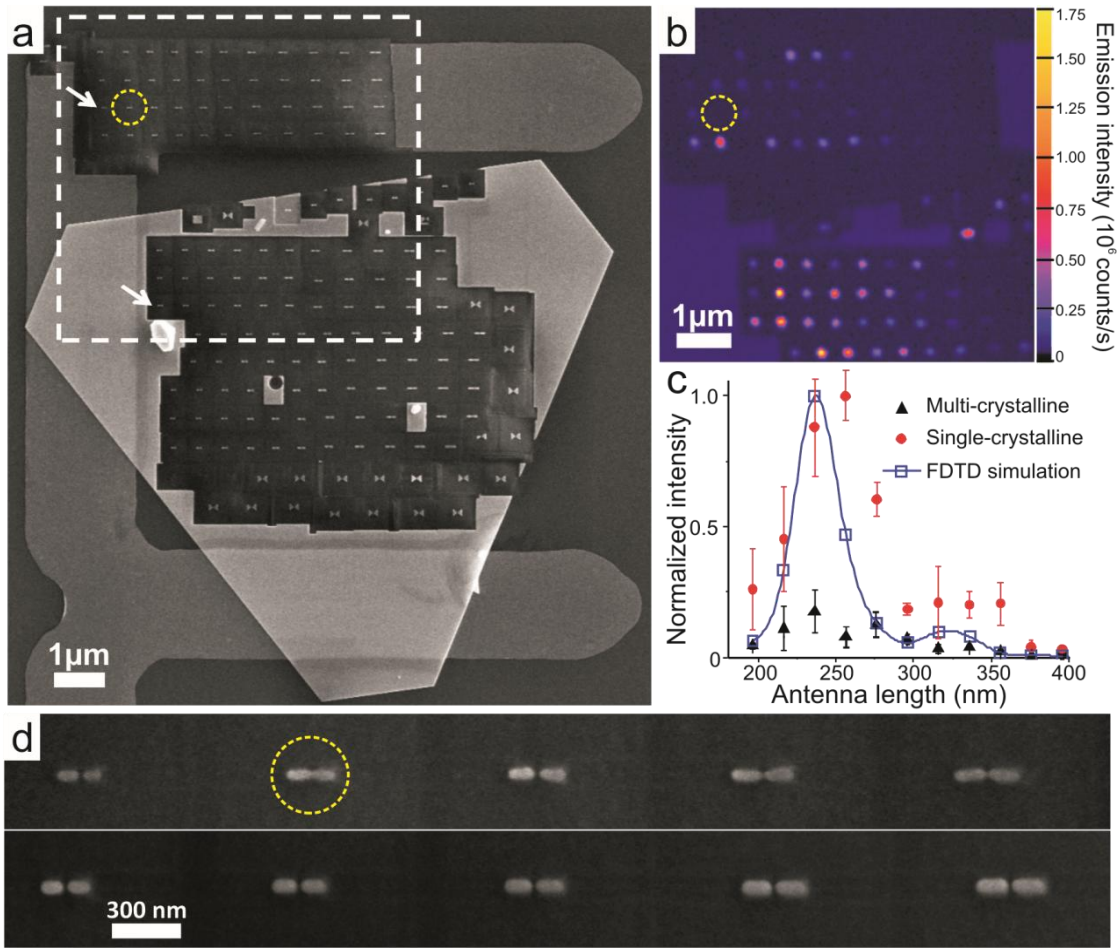

# Supplementary Information

Atomically flat single-crystalline gold nanostructures for plasmonic nanocircuitry

Jer-Shing Huang[1,*], Victor Callegari[2], Peter Geisler[1], Christoph Brüning[1], Johannes Kern[1], Jord C. Prangsma[1], Xiaofei Wu[1], Thorsten Feichtner[1], Johannes Ziegler[1], Pia Weinmann[3], Martin Kamp[3], Alfred Forchel[3], Paolo Biagioni[4], Urs Sennhauser[2] & Bert Hecht[1,†]

*1. Nano-Optics & Biophotonics Group, Experimentelle Physik 5, Physikalisches Institut, Wilhelm-Conrad-Röntgen-Center for Complex Material Systems, Universität Würzburg, Am Hubland, D-97074 Würzburg, Germany*

*2. EMPA, Swiss Federal Laboratories for Materials Testing and Research, Electronics/Metrology Laboratory, CH-8600 Dübendorf, Switzerland*

*3. Technische Physik, Physikalisches Institut, Wilhelm-Conrad-Röntgen-Center for Complex Material Systems, Universität Würzburg, Am Hubland, D-97074 Würzburg, Germany*

*4. CNISM - Dipartimento di Fisica, Politecnico di Milano, Piazza Leonardo da Vinci 32, 20133 Milano, Italy*

**\*** jshuang@mx.nthu.edu.tw

(J.-S.H. current address: Department of Chemistry, National Tsing Hua University, Hsinchu 30013, Taiwan)

[†] hecht@physik.uni-wuerzburg.de

# 1. Preparation and characterization of self-assembled single-crystalline gold microflakes

## 1.1. Scanning electron microscopy (SEM)

Single-crystalline gold flakes are prepared following the procedure described in ref. 40. The flake suspension is sonicated for 5 minutes right before being drop-casted on the ITO glass substrate. After the solvent dries out, flakes with sufficiently large area and homogeneous contact with the substrate are pre-selected using wide-field optical microscope with a low-magnification air objective (20x). The sample is then transferred into the vacuum chamber for SEM and focused-ion beam (FIB) milling. Pre-selected flakes are double checked with SEM, where the homogeneity of image brightness is used as a positive indicator. Multi-crystalline structures are fabricated on EBL fabricated marker structures nearby the selected flakes to ensure identical local environment and identical FIB focusing conditions. All the structures in this work are milled with Ga-ions using an acceleration voltage of 30 kV and a current of 1.5 pA.

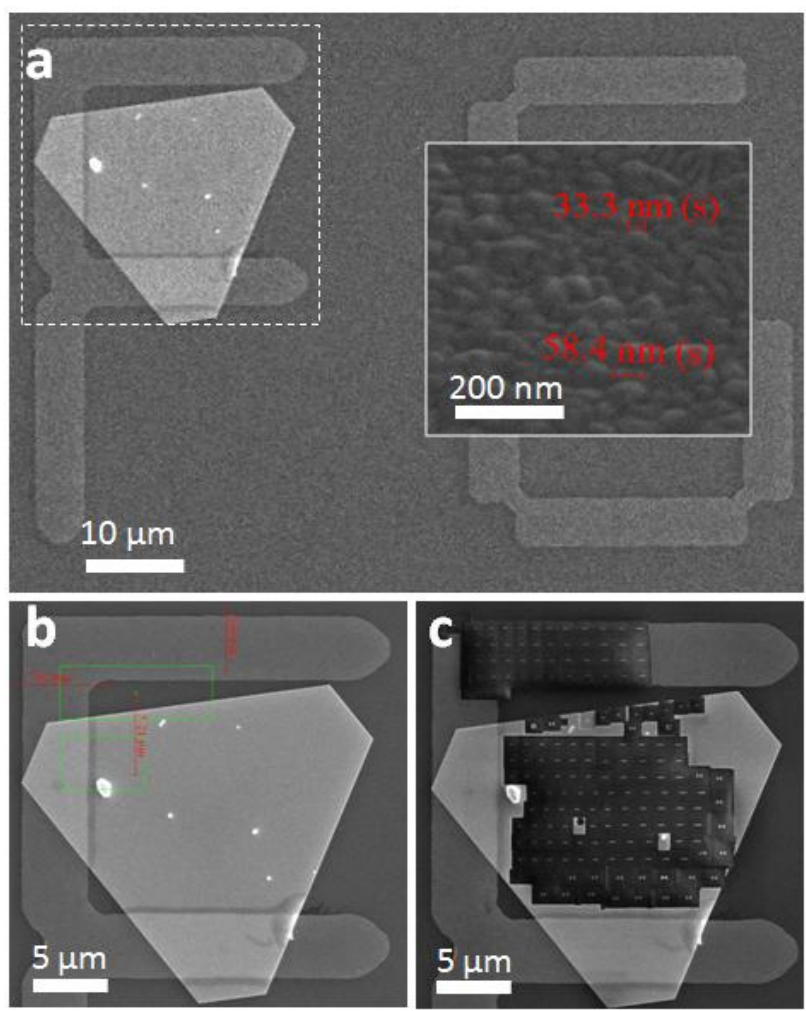

**Figure S1 | SEM image of an area containing chemically synthesized single- and vapor-deposited multi-crystalline gold films.** (a) Overview of the area of interest including both parts of the multi-crystalline marker structure and the single-crystalline flake. The inset shows a zoomed-in image of the typical surface of vapor-deposited multi-crystalline gold film consisting of randomly orientated grains. (b) Zoomed-in SEM image of the dashed area in (a) before FIB milling. (c) Same area after FIB milling.

Although the same FIB patterns are used to fabricate the structures, the ion-beam dose is optimized for the multi-crystalline and single-crystalline areas, respectively, so that the finest features, e.g. the gap size of the nanoantennas or the width of a linear cut, were comparable. Figure S1 shows the SEM image of the area of interest including flake and marker stucture (letter F) before and after FIB milling.

### 1.2. Atomic force microscopy (AFM)

Figure S2a-c display AFM topographies of the structures presented in Figure 4 in the main text, while Figure S2d displays line cuts through single- and multi-crystalline antenna arrays as well as two additional line cuts addressing the thickness of the metal films and the cutting depth into the ITO substrate. The baseline (height = 0 nm) is set to the level of the unpatterned ITO surface. Before writing the nanoantennas, the chosen flake (thickness about 65 nm, see L2 in Figure S2d) is homogeneously thinned by a large area low-dose FIB to reduce the thickness to that of the multi-crystalline film (25-30 nm, see L1 profile in Figure S2d). Consequent FIB milling of the marker structure and the thinned flake yields single- and multi-crystalline antennas with a narrow height distribution as shown in Figure 2d. The topography contrast of 55-60 nm around nanoantennas is the sum of the respective antenna height (30-35 nm) and the ITO milling depth (20- 25 nm, see L1 and L2 in Figure 4d).

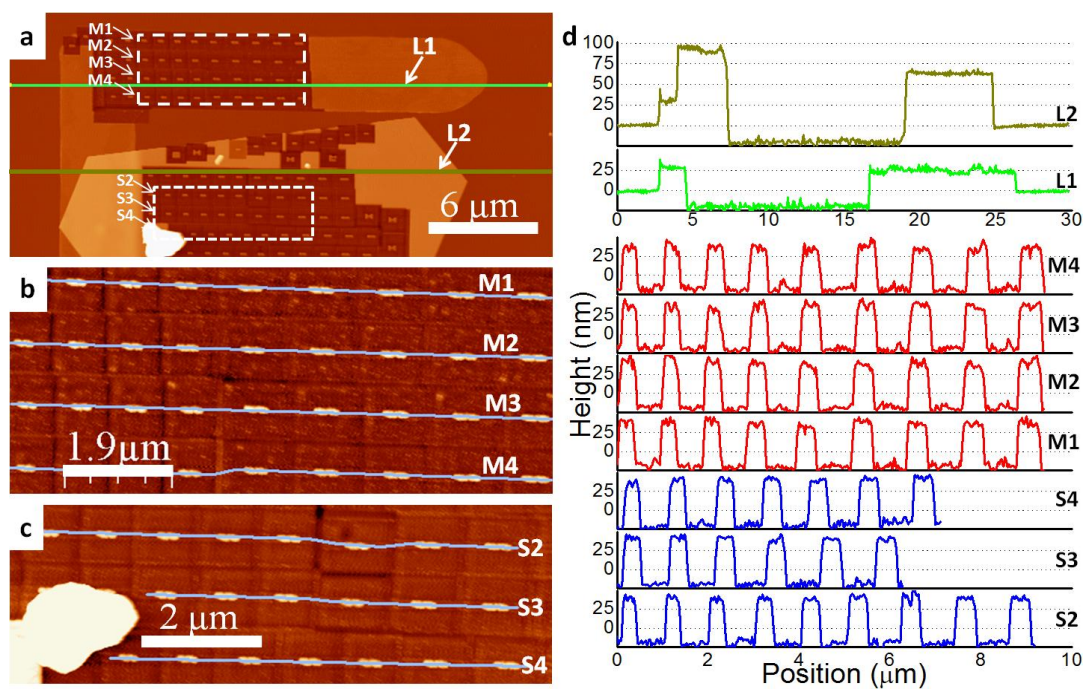

**Figure S2 | Atomic force microscope (AFM) images and line cuts through single- and multi-crystalline antenna arrays** (a) AFM image of the multi- (dashed rectangle) and single-crystalline (dotted rectangle) nanoantenna arrays together with marker structures, as shown in Figure S1c (b) Zoomed-in AFM image of the multi-crystalline antenna arrays (c) Zoomed-in AFM image of the single-crystalline antenna arrays (d) Line cuts of the antenna arrays recorded along the lines indicated in Figure S2b and Figure S2c and two topography profiles recorded along the lines marked as L1, and L2 in Figure S2a. The big particle near antenna array S4 (Figure S2a) has been moved due to the AFM measurement, as can be seen by comparing its position in Figure S2a and Figure S2c.

### 1.3. Optimization of the flake preparation procedure

In order to maximize the size of the flakes and therefore the area available for FIB milling, we have decreased the temperature of the chemical reaction to 60˚C and increased the reaction time to more than 12 hours. Figure S3 shows a histogram illustrating the temperature depencence of the flake size as obtained by evaluating optical images of flakes deposited on substrates as described above. While the average flake size increases with decreasing reaction temperature, the average flake thickness remains smaller than 80 nm which facilitates the fabrication.

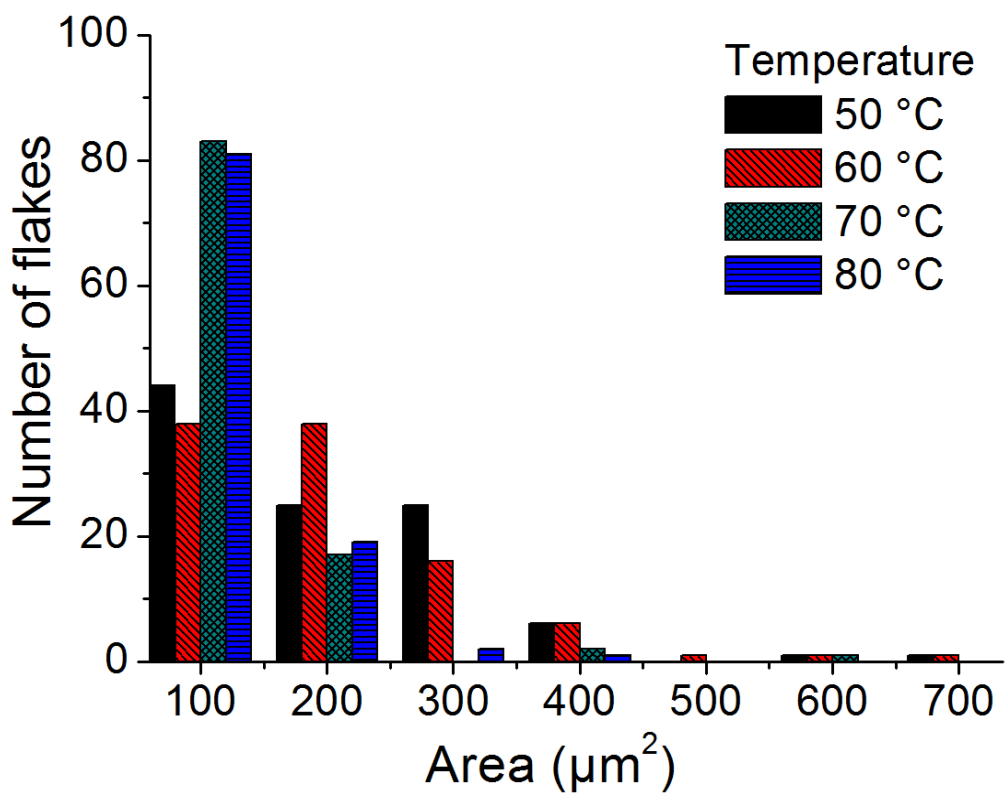

**Figure S3 | Flake area distribution with respect to the reaction temperature.** The histogram is obtained by measuring the area of 100 randomly sampled flakes from each sample suspension.

**1.4. Characterization of the gold flake surface roughness**

We have performed atomic force microscopy (AFM) and confirm that the surface roughness of the flake is smaller than 1 nm over an area of 1 $\mu m^2$. A typical AFM image with corresponding topography histogram is shown in Figure S4. The fluctuation of the surface height is mainly due to the surfactant and other contaminations from the solvent. In addition, extremely low yield of visible two-photon photoluminescence (TPPL) signal from the single-crystalline flakes can also serve as an evidence for the ultrasmooth flake surface. As shown in Figure S5, even for 500 µW focused excitation (830 nm, 1 ps, N.A.=1.4), the flat flake shows no significant spectrally dispersed TPPL as compared to the multi-crystalline marker area and nanoantennas, which is in agreement with previous studies [refs. 49, S1].

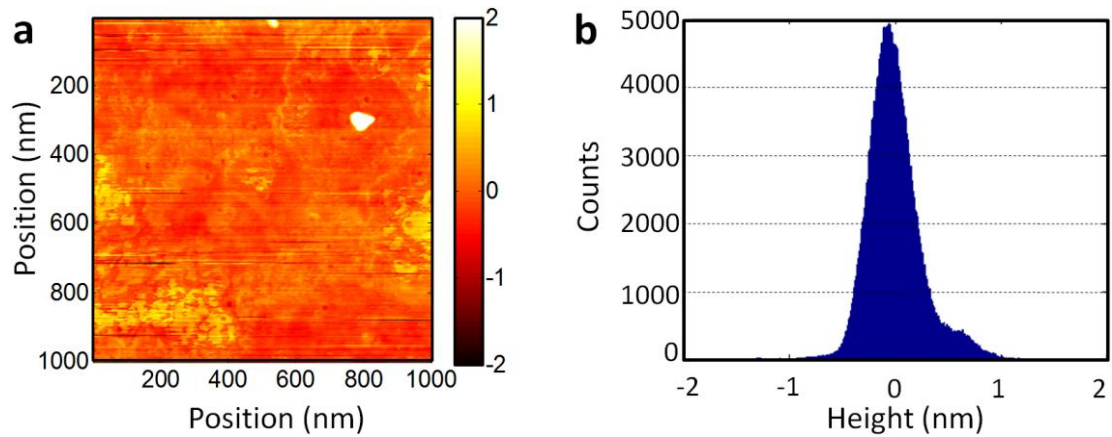

**Figure S4 | AFM images of a single-crystalline flake surface.** (a) Typical AFM image of the surface topography of a single-crystalline flake. Scale unit: nm (b) Distribution of the surface height over a 1 µm$^2$ area. The surface roughness is smaller than 1 nm.

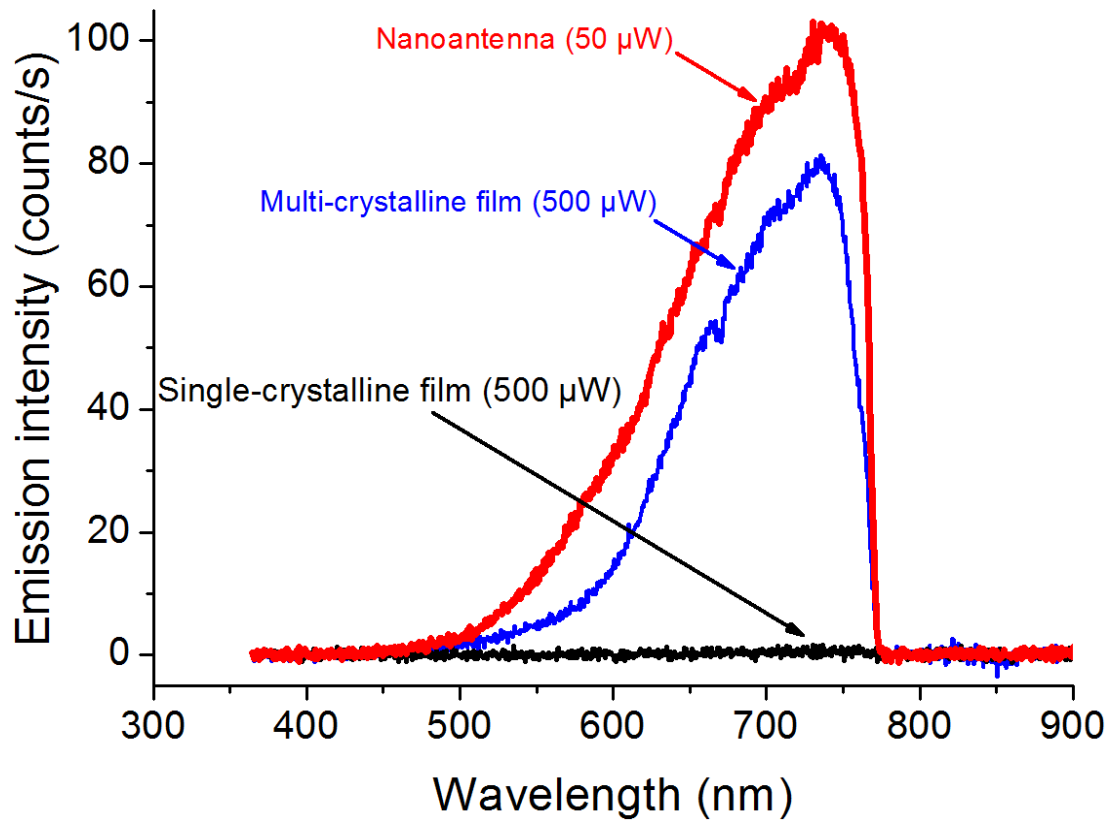

**Figure S5 | Two-photon photoluminescence (TPPL) spectra of a resonant nanoantenna as well as single- and multi-crystalline gold films.** Due to the surface roughness, unstructured multi-crystalline gold marker (blue thin line) exhibits much stronger TPPL signal compared to the unstructured single-crystalline flake (black medium line). However, when the single-crystalline gold film is shaped into a resonance linear dipole nanoantenna (50 nm wide, 30 nm high and 236 nm long with 16 nm feedgap, red thick line), the TPPL signal is greatly enhanced. Note that the excitation polarization for the nanoantenna curve is along the antenna long axis and the excitation power is 50 µW (830 nm, 1 ps, N.A. = 1.4).

## 2. Two-photon photoluminescence confocal microscopy

The distance between adjacent antennas is larger than 700 nm to minimize crosstalk and to ensure that only one antenna is illuminated by the focal spot (FWHM = 350 nm). The ultrashort pulses from a mode-locked Ti:sapphire laser (center wavelength = 828 nm, 80 fs, 80 MHz, average power: 50 - 500 µW, Time-Bandwidth Products, Tiger) are coupled into 1.5 m of optical fiber to get stretched to 1 ps, which avoids possible damage of the nanoantennas without decreasing the TPPL signal. The linear polarized fiber output is then collimated, passes a dichroic mirror (DCXR770, Chroma Technology Inc.) and is focused through the cover glass onto the antenna array using an oil immersion microscope objective (Plan-Apochromat 100x, Oil, N.A. = 1.4, Nikon). The direction of linear polarization of the beam is adjusted using a $\lambda/2$-plate. The photoluminescence signal is collected by the same objective and is reflected by the dichroic mirror. Laser scattering and possible second harmonic signals are rejected by a holographic notch filter (O.D. > 6.0 at 830 nm, Kaiser Optical System, Inc.) and a bandpass filter (transmission window: 450-750 nm, D600/300, Chroma Technology Inc.) in front of the photon detector (SPCM-AQR 14, Perkin-Elmer). A typical emission spectrum of a resonant nanoantenna is shown in Figure S5. The quadratic dependence of the recorded integrated signal on the excitation power confirms that the visible TPPL of gold dominates the antenna emission (Figure S6). Figure S7 shows the dependence of the TPPL signal intensity on the excitation polarization which verifies that the longitudinal antenna resonance is excited.

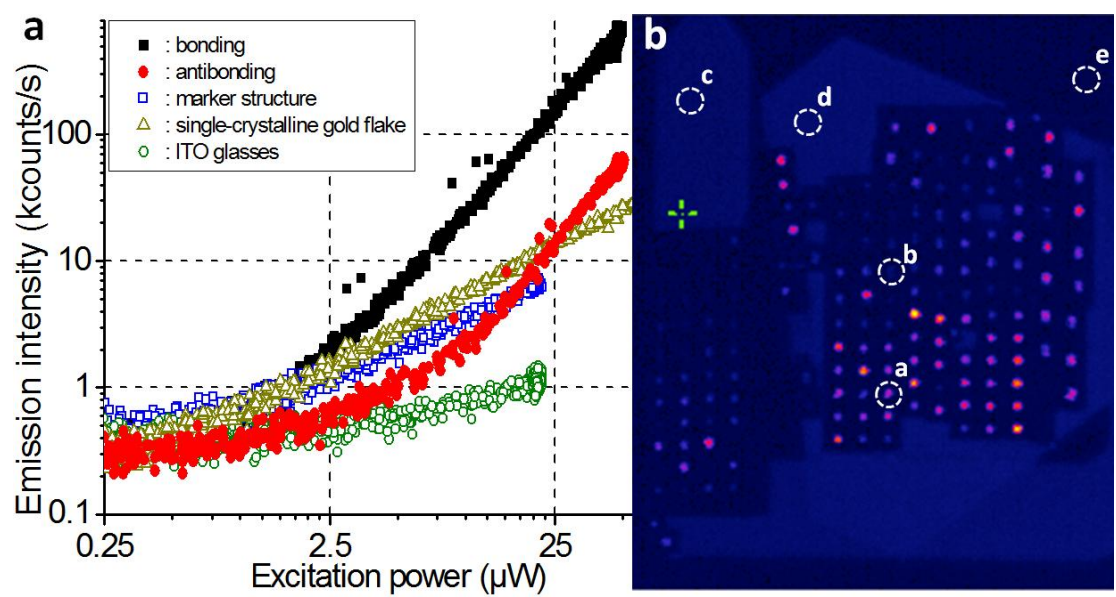

**Figure S6 | Intensity dependence of visible TPPL signals on the average excitation power** (a) Emission intensity as a function of the excitation power obtained from the areas marked with the dashed circles in the emission map (b). The emission signals from antennas with bonding (black solid squares) and antibonding (red solid dots) resonance show quadratic dependence on the excitation power while the scattering from the multi-crystalline gold marker structure (blue open square), single-crystalline gold flake (yellow open triangle) and bare ITO glass area (green open circle) show very weak scattering with linear dependence on the excitation power.

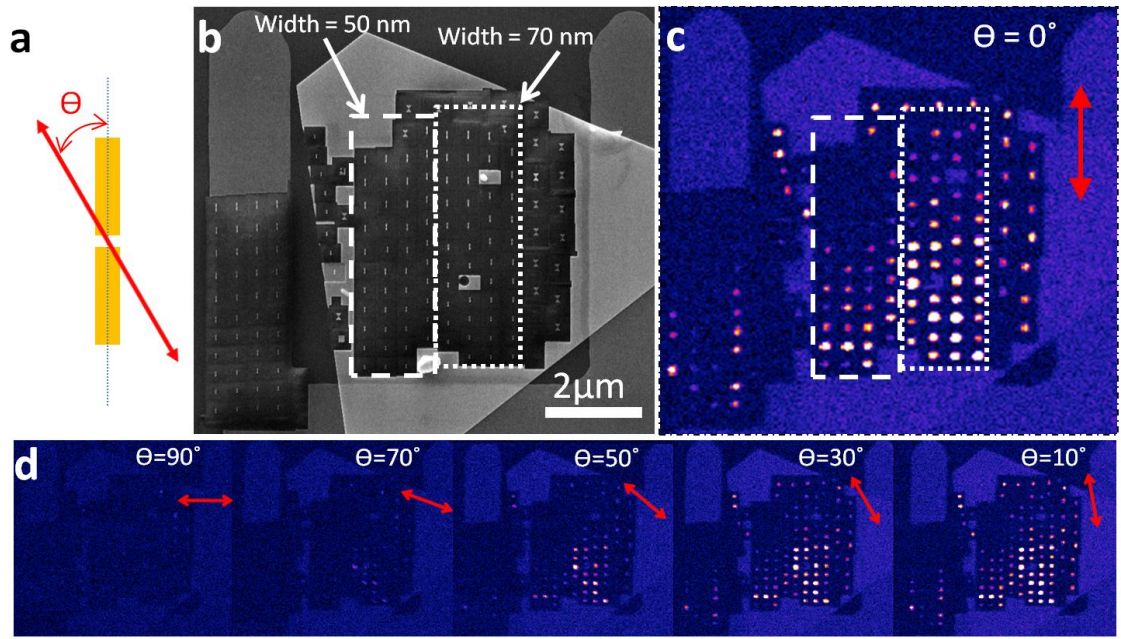

**Figure S7| SEM image and TPPL maps of the whole antenna array obtained with different excitation polarizations.** (a) Schematic diagram of the linear excitation polarization (red double arrows). Θ is defined as the angle between the excitation polarization and the antenna's long axis; (b) SEM image of the fabricated area including arrays of antennas with nominal width of 50 nm (dashed rectangle) and 70 nm (dotted rectangle) ; (c) TPPL map of the corresponding area shown in (b) with longitudinal excitation (Θ = 0°); (d) TPPL maps for various excitation polarizations. Same intensity scale for all TPPL maps.

## 3. Numerical simulations

The resonance of the fabricated bowtie nanoantennas in Figure 2 is simulated using the nominal dimensions used in the FIB milling. The gold bowtie nanoantennas are placed on top of an ITO layer. Since the antenna response is sensitive to the local index of refraction, also the geometry changes of the substrate due to the FIB milling are taken into account. Figure S8a-c shows the top view and cross section of the simulated structure. The dielectric function of gold is described by an analytical model [ref. S2] which fits the experimental data [ref. 60], while the dielectric function of the sputtered ITO layer is based on experimental data [ref. S3]. A multi-coefficient model [ref. S4] is then used to fit the dielectric function within the frequency window of interest to gain speed in the simulation. A uniform mesh volume with discretization of 1 nm$^3$ covers the whole antenna and all the boundaries of the simulation box are set to be at least 700 nm away from the antenna to avoid spurious absorption of the antenna near fields. Corners of the antenna arms are rounded in the simulations using cylinders with 10 nm diameter. The source is set to have the same spectral width as the laser used in the experiment (821 - 835 nm) and is focused onto the gold/ITO interface at the feedgap center using a thin lens (N.A. = 1.4) focusing approximated by a superposition of 200 plane waves. The impulse spectrum is recorded at the center of the feedgap and normalized to the source spectrum. Figure S8d shows the simulated spectra of the bowtie nanoantenna sketched in Figure S8a-c, in which the laser spectrum used in the experiments indicated by the shaded area overlaps only partially with the resonance. Slightly different spectra are obtained for different microscopic conditions at the feedgap as indicated in Figure S8d.

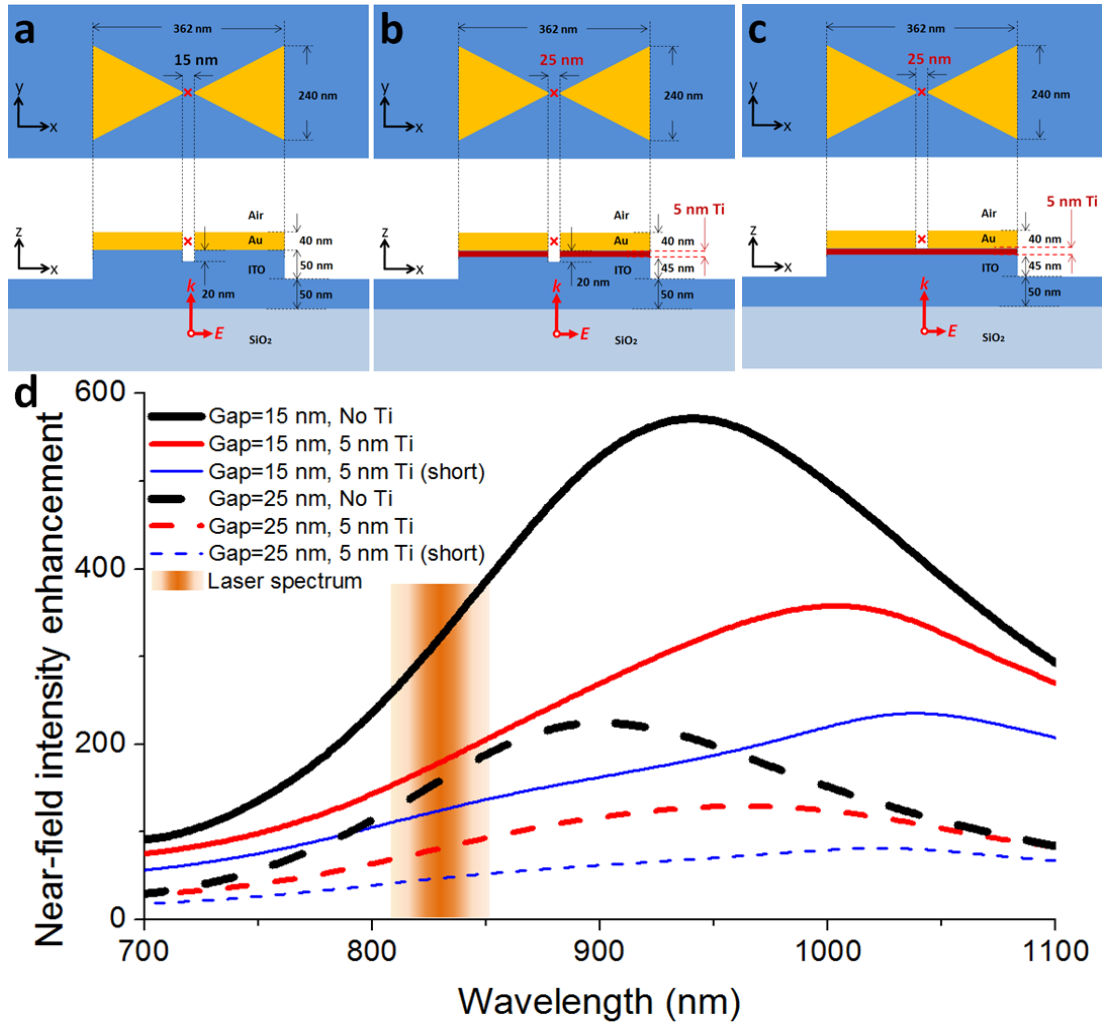

**Figure S8 | Schematic diagram of the simulated structures and the impulse spectra** (a) Top view and cross section of a gold bowtie antenna with (a) 15nm gap on ITO substrate, (b) a 25 nm gap and 5 nm Ti adhesion layer between gold and ITO substrate cut through in the gap and (c) a 25 nm gap and 5 nm Ti adhesion layer between gold and ITO substrate not cut through in the gap. The excitation source (symbolized by the red arrows) is focused onto the feedgap center. Near-field intensity spectra are recorded at the feed gap center (red crosses). (d) The simulated spectra of the bowtie nanoantennas sketched in (a, thick black), (b, medium red) and (c, thin blue) with gap of 15 (solid line) and 25 nm (dashed line). The shaded area indicates the real source spectral range, which does not hit the resonance peak but partially overlaps with the broad resonance peak.

## 4. Ga-ion implantation by FIB milling

In order to check for Ga implantation into the gold flake due to FIB milling, energy-dispersive X-ray (EDX) analysis was performed at the patterned edge of the nanostructure with a FEI Titan 80-300 transmission electron microscope (TEM) equipped with a retractable X-ray detector (EDAX). The sample was prepared by deposition of single-crystalline gold flakes on a thin carbon film. A suitable flake was identified and patterned by a focused Ga ion beam in a dual-beam system (FEI Helios nanolab). The sample was then transferred to the TEM, which was operated at 300 kV acceleration voltage and a beam current of 1 nA. Under these conditions, the beam diameter is around 1 nm. The sample was tilted by 10 degree towards the detector in order to increase the count rate. After imaging the sample in scanning mode using a high-angle annular dark field detector, the beam was positioned at the edge of a cut in the gold flake and the X-ray spectrum shown in Figure S9a was recorded with a width of 5eV per channel and 64 seconds integration time. The energy resolution of the detector was 129 eV at 5.9 keV X-ray energy. The Fe, Co and Cu lines that are visible in the spectrum are due to electrons that are scattered by the sample and hit the TEM column, the pole piece of the objective lens or the copper support grid of the carbon film.

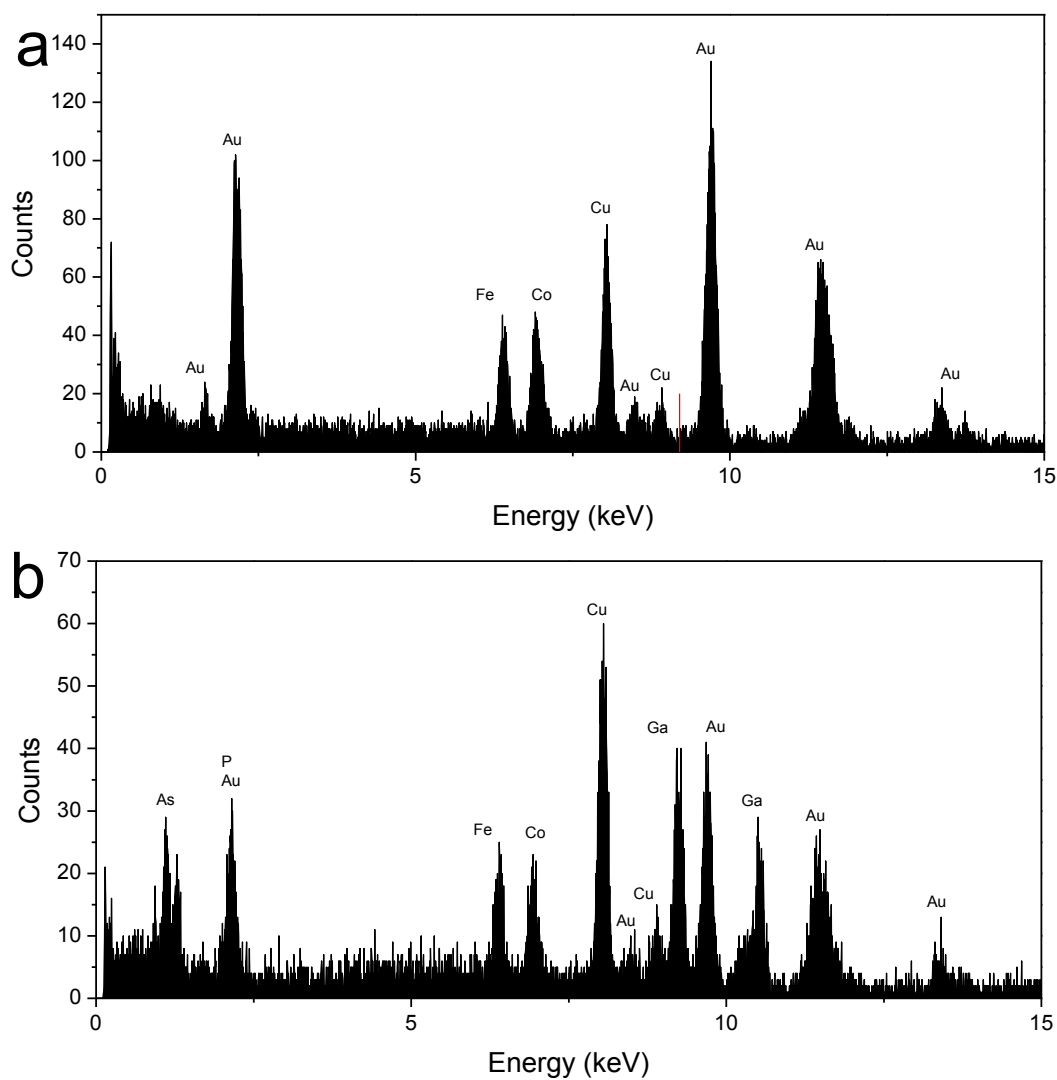

**Figure S9 | EDX spectrum recorded at the FIB milled edge of a single crystalline gold flake** (a) EDX spectrum recorded at the patterned edge of a single-crystalline Au flake (b) Reference EDX spectrum from a gold coated GaAsP nanowire. The strongest Ga ion peak, indicated with a red line in (a), is at the noise level corresponding to a concentration less than 1%. Fe, Co and Cu lines are due to electron scattering which hits the TEM column, pole piece of the objective lens and the sample holder.

In order to quantify the upper limit of Ga at the edge of the gold flake, we recorded the X-ray spectrum of a reference sample (gold coated GaAsP nanowire) that was also deposited on a carbon film (see Figure S9b). The spectrum of the reference sample was analyzed using the TIA (TEM Imaging and Analysis) software package from FEI. The compositional analysis of the reference sample (excluding the Fe, Co and Cu peaks) was then related to the count rates in the energy range where the Ga peak shows up (9-9.4 keV), leading to an estimated upper bound of 1% for the Ga concentration.

## 5. White-light scattering spectra

In addition to the AFM measurements, which prove the fact that the single- and multi-crystalline antenna structures have very similar dimensions, we have also recorded white-light scattering spectra of all antennas shown in Figure 4 to provide an independent proof of the fact that the fabricated antenna arrays include antennas that are on resonance with the 830 nm excitation light. As shown in Figure S10, the peak positions and widths of selected (resonance in the visible/near infrared) single- and multi-crystalline antennas with same nominal dimensions are very close to each other, indicating very similar dimensions, in accordance with the AFM results. Compared to the simulated spectra (black dashed lines in Figure S10), the experimental resonances seem blue-shifted. This can be attributed to the possibly larger effective gap size in the fabricated structures which results in blue shifted spectra. We would like to note that the agreement with simulated spectra can still considered to be very good, since only small changes in the dimensions of the simulated structures would be sufficient to improve the agreement between experiment and simulations.

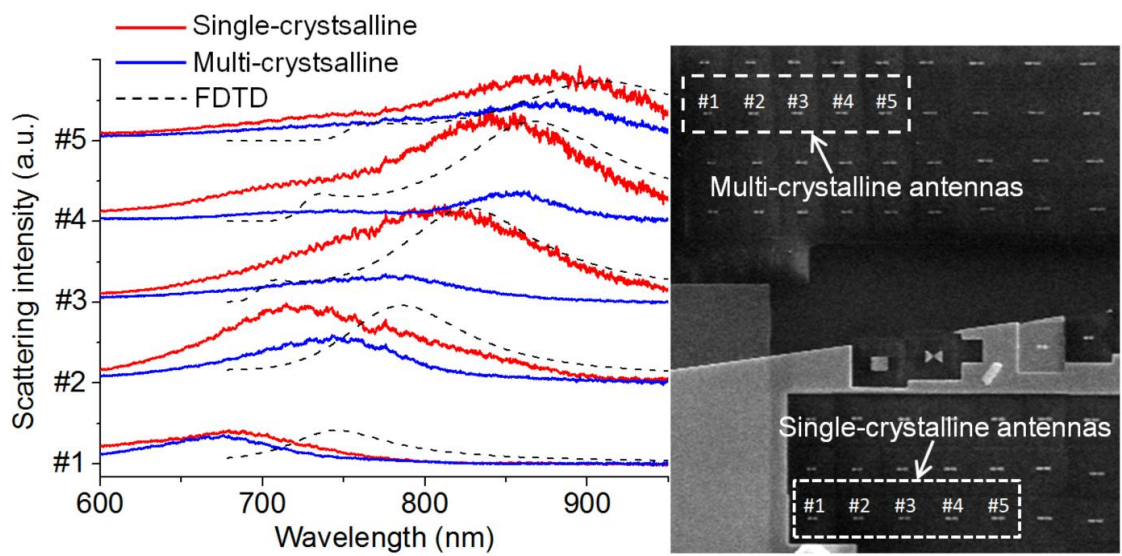

**Figure S10 | White-light scattering spectra and simulated near-field spectra of the nanoantennas (Plotted with increasing offset for clarity).** White-light scattering spectra (see Methods) for single- (red) and multi-crystalline antennas (blue) together with simulated near-field spectra (black) obtained using a FDTD simulation as described in ref. 52 (for details see Methods, Numerical simulations) for antennas as labeled in the SEM image (right panel). The simulated spectra are normalized to the amplitude of the single-crystalline antenna spectra.

In Figure S10, it is also clearly seen that the single-crystalline nanoantennas show much higher scattering intensity than multi-crystalline ones while the spectral line width is more or less comparable. Such higher scattering intensity is in accordance with the stronger TPPL signal and supports one of our main findings, which is that the single-crystalline nanoantennas exhibit superior optical properties. While the comparable dimensions, i.e. height, width, length and gap size, of antennas in both arrays result in very close resonance peak position, the imperfections in the nanoantenna shape, including bumps, tilted arms and even shorted gaps (see Figure 4d) lead to worse-defined multi-crystalline antennas with lower scattering and distorted spectra. Therefore, compared to multi-crystalline antennas, single-crystalline antennas may have lower non-radiative decay rate due to the single crystallinity [ref. S5] but higher radiative decay rate due to a better defined shape. However, due to the additional geometrical factors, the observed line width and scattering amplitude cannot be intuitively understood since different contributions to the total decay rate cannot be easily disentangled. A quantitative comparison of the observed line widths is interesting but not straight forward and needs further systematic study.

## Supplementary Reference